# The nonrelativistic limit of the second order S-matrix element for the elastic scattering of photons by K-shell bound electrons

A. Costescu [a)], S. Spanulescu [a), b)], C. Stoica[a)]

[a)] Department of Physics, University of Bucharest, MG11, Bucharest-Magurele 76900, Romania

[b)] Department of Physics, Hyperion University of Bucharest, Postal code 030629, Bucharest, Romania

## Abstract

In this paper, the right expressions of the fully nonrelativistic K-shell Rayleigh scattering amplitudes and cross-sections are obtained by using the Coulomb Green function method. Our analytical result has no spurious poles as the old nonrelativistic result with retardation [M. Gavrila and A. Costescu Phys. Rev. A **2**, 1752 (1970)] presents.

Taking into account the exact expression of the second order S-matrix element in the case of the elastic scattering of photons by K-shell bound electrons we obtain its nonrelativistic limit (valid for photon energies $\omega$ up to $\alpha Z m$, for intermediate Z), by removing in a consistent and appropriate way the higher order relativistic terms in $\alpha Z$ and $\omega/m$.

We obtain the imaginary part of the Rayleigh amplitudes in terms of elementary functions. Thereby the exact nonrelativistic total photoeffect cross-section is obtained *via* the optical theorem.

Comparing the predictions given by our formulae with the full relativistic numerical calculations of Kissel *et al* [Phys. Rev. A **22**, 1970 (1980)], and with experimental results, a fairly good agreement within 10% is found for the angular distribution of Rayleigh scattering for photon energies up to $\alpha Z m$, for intermediate Z, and up to twice the photoeffect threshold energy for high Z elements. Considering the relativistic coulombian location for the photoeffect threshold very good predictions are found below the first resonance, even in its vicinity.

We present our numerical results for Rayleigh angular distribution and the photoeffect total cross-section, for various photon energies, scattering angles and Z, compared with other theoretical and experimental results.

## I. INTRODUCTION

As it was previously shown by Costescu *et al* [3], in the case of the photon elastic scattering from K-Shell electrons, the well-known nonrelativistic Kramers-Heisenberg-Waller (KHW) matrix element presents spurious poles located at $\omega = 2m$, far enough from the K-shell photo-effect threshold $\omega_{th}$.

The KHW matrix element for Rayleigh scattering of an initial photon with momentum $\vec{k}_1 = \omega \vec{v}_1$ and polarization vector $\vec{s}_1$ and a final photon with momentum $\vec{k}_2 = \omega \vec{v}_2$ and polarization vector $\vec{s}_2$ is [1]:

$$\mathcal{M}_{RET}^{NR} = M_{RET}(\omega,\theta)(\vec{s}_1 \vec{s}_2) + N_{RET}(\omega,\theta)(\vec{s}_1 \vec{v}_2)(\vec{s}_2 \vec{v}_1), \qquad (1.1)$$

where $\theta$ is the photon scattering angle and

$$M_{RET}(\omega,\theta) = [\vartheta - P_{RET}(\Omega_1,\theta) - P_{RET}(\Omega_2,\theta)], \qquad (1.2)$$

$$N_{RET}(\omega,\theta) = -[Q_{RET}(\Omega_1,\theta) + Q_{RET}(\Omega_2,\theta)], \qquad (1.3)$$

$$P_{RET}(\Omega,\theta) = \frac{128\lambda^5 X^3}{d_{GC}^4(\Omega)} \frac{F_1(2-\tau;2,2;3-\tau;x_1,x_2)}{2-\tau}, \qquad (1.4)$$

$$Q_{RET}(\Omega,\theta) = \frac{2048\lambda^5 X^5 \omega^2}{d_{GC}^6(\Omega)} \frac{F_1(3-\tau;3,3;4-\tau;x_1,x_2)}{3-\tau}, \qquad (1.5)$$

where $\lambda = \alpha Z m$,

$$\Omega_1 = -\lambda^2/2m + \omega, \quad \Omega_2 = -\lambda^2/2m - \omega, \qquad (1.6)$$

$$X_j^2 = -2m\Omega_j, \tau_j = \frac{\lambda}{X_j}, \quad \mathrm{Re}[X_j] > 0, \, j=1,2$$

$$(1.7)$$

The function $\vartheta = \left[1 + \frac{1}{\alpha^2 Z^2}\left(\frac{\omega}{m}\right)^2 \sin^2\frac{\theta}{2}\right]^{-2}$ is the atomic form factor, while $F_1(a;b_1,b_2;c;x_1,x_2)$ is the Appell hypergeometric function of four parameters and two complex variables given by the relationships:

$$x_1 x_2 = p = \left[\frac{d_{GC}^*(\Omega)}{d_{GC}(\Omega)}\right]^2 \equiv \xi^2, \qquad (1.8)$$

$$x_1 + x_2 = s = 2\xi - \frac{16 X^2 \omega^2 \sin^2\frac{\theta}{2}}{d_{GC}^2(\Omega)}. \qquad (1.9)$$

with

$$d_{GC}(\Omega) = (X + \lambda)^2 + \omega^2 \quad , \quad d_{GC}^*(\Omega) = (X - \lambda)^2 + \omega^2. \qquad (1.10)$$

According to the optical theorem the imaginary part of the Rayleigh amplitude for forward scattering allows to get the total photoeffect cross-section *i. e.* the well-known Fischer formula [4]:

$$\sigma_{ph} = \frac{32\pi^2}{3\alpha}(\alpha Z)^6 \left(\frac{m}{\omega}\right)^4 \frac{e^{-|\tau_1|\chi(k_{NR},Z)}}{1 - e^{-2\pi|\tau_1|}} \frac{1}{\left[(1-\omega/2m)^2 + \alpha^2 Z^2\right]^2} r_0^2 \qquad (1.11)$$

where $\chi(k_{NR},Z) = \arctan\frac{2(k_{NR}-1)^{1/2}}{2 - k_{NR} + (\omega/\alpha Z m)^2}$, $k_{NR} = \omega/I_B$ and $I_B = \frac{\alpha^2 Z^2}{2} m$ is the nonrelativistic photoeffect threshold.

Inspecting equations (1.4), (1.5) and (1.12), we notice that above the photoeffect threshold we also get:

$$|d(\Omega_1)|^2 = 4m^2\omega^2\left[(1-\omega/2m)^2 + \alpha^2 Z^2\right]. \qquad (1.12)$$

It is obvious that when the photon energy $\omega$ is greater than 100 keV the KHW result strongly overestimates the Rayleigh cross-section, because the spurious poles $\omega = 2m \pm i2\alpha Zm$ are at least of the 4th order in the amplitude i.e. of the 8th order in the cross-section, as well as the photoeffect cross-section. We notice that for low and intermediate Z values and photon energies up to three times the photoeffect threshold energy, the modulus $|d(\Omega_1)|$ of the denominator is close to the value $2m\omega$. In the next section we prove that $|d(\Omega_1)| = 2m\omega$ is the right nonrelativistic limit of the denominator that explains why the nonrelativistic result obtained by Gavrila and Costescu [1] in KHW approach making use of the coulombian Green function of Schwinger [5] gives fairly good predictions above the photoeffect threshold up to 20-25 keV for low and up to 90-100 keV for intermediate Z values [3], [6].

Recently, it has been proven [7] that in the case of Compton scattering on K-shell bound electrons the nonrelativistic KHW matrix element leads to inadequate expressions for the nonrelativistic Compton amplitudes when the sum over the complete set of positive-energy intermediate states is removed by using the nonrelativistic coulombian Green function and its integral representation. Indeed, in this way the nonrelativistic kinematics would be taken into account from the very beginning of the analytical calculation, so that important relativistic kinematics terms in $\omega^2$, which would in fact exactly cancel some terms due to multipoles, should be omitted. This is why the spurious kinematics poles occur in the KHW amplitudes for any two photon process.

The right way to obtain the nonrelativistic amplitudes for a two photon atomic process is to consider the second order S-matrix element involving the sum over the complete set of intermediate states $|n\rangle$ (solutions of Dirac equation with Coulomb field) of energies $E_n$, corresponding to both positive and negative energy states and replacing this sum with the coulombian Green function of Dirac equation given by Hostler and Pratt [8,9]. In this way the relativistic kinematics is taken into account when performing the integrals involved in the matrix element. Only after performing all needed integrals the nonrelativistic limit of the second order S-matrix elements should be considered and the right nonrelativistic amplitudes are obtained without any spurious singularity.

In the specific case of the elastic scattering of photons from K-shell bound electrons, both final and initial states are described by Dirac spinors which have significant values only in a region located near the nucleus at a distance of the order of the ion Bohr radius $a_0/Z$ where $a_0$ =0.53 Å. This shows that small distances contribute to the matrix element much more in the case of Rayleigh scattering than in the case of Compton scattering where the final state belong to continuum spectrum and the overlapping at small distances with the ground state Dirac spinors is less important. As a consequence, the validity of the nonrelativistic result has to come to an end at significantly smaller energies in the case of Rayleigh scattering than in the case of Compton scattering.

In the next section we present our analytical calculations in order to get the right nonrelativistic limit of the S-matrix element (NRLSM) for Rayleigh scattering on ground state electrons.

We also present numerical comparisons with relativistic S-matrix calculations [2] that confirm the validity of our NRLSM over a significantly larger energy range than the old nonrelativistic result of Gavrila and Costescu (NRGC) [1], the NRLSM being valid above the photoeffect threshold for photon energies up to 200 keV both for intermediate and high Z elements.

Accordingly, the nonrelativistic limit of the imaginary part of the Rayleigh amplitudes is correctly obtained, particularly the imaginary part of the forward scattering amplitude, allowing to get the exact expression for the total photoelectric K-shell cross-section, as it no longer contains spurious singularities.

The aim of the present paper is to obtain the right angular distribution of the Rayleygh scattering cross-section on bound electrons in the nonrelativistic limit, without spurious singularities, and to evaluate the domain of it's validity.

In the literature is known the numerical full relativistic (RMP) calculations of Kissel *et al* [2] for the Rayleigh scattering cross-section on bound electrons, which makes the direct numerical calculations starting from the S-matrix second-order element. These results allow us to compare our predictions with the full relativistic ones. Also, below the first resonance we compare our results with the relativistic dipole approximation for forward Rayleigh scattering of Florescu et al [10].

## II. The right nonrelativistic limit of the Rayleigh second order S-matrix element.

There are two Feynman diagrams describing the second-order amplitudes of the Rayleigh scattering process (Fig. 1).

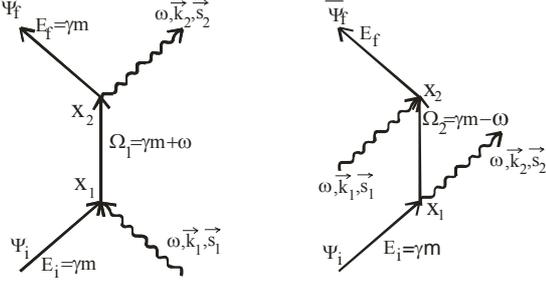

Fig.1 Feynman diagrams for Rayleigh scattering

The S-matrix element for Rayleigh scattering of a photon energy $\omega$, momentum $\vec{k}_1 = \omega \vec{v}_1$ and polarization $\vec{s}_1$ from a K-shell bound electron of polarization $\mu_i$ is given by the expression:

$$\mathcal{M}^R_{\mu_f \mu_i} = \mathcal{M}_{\mu_f \mu_i}(\Omega_1) + \mathcal{M}_{\mu_f \mu_i}(\Omega_2) \qquad (2.1)$$

where

$$\mathcal{M}_{\mu_f \mu_i}(\Omega_1) = -m \langle f | \vec{\alpha}\,\vec{s}_2 e^{-i\vec{k}_2 \vec{r}_2} G(\vec{r}_2,\vec{r}_1;\Omega_1) \vec{\alpha}\,\vec{s}_1 e^{i\vec{k}_1 \vec{r}_1} | i \rangle \qquad (2.2a)$$

$$\mathcal{M}_{\mu_f \mu_i}(\Omega_2) = -m \langle f | \vec{\alpha}\,\vec{s}_1 e^{i\vec{k}_1 \vec{r}_2} G(\vec{r}_2,\vec{r}_1;\Omega_2) \vec{\alpha}\,\vec{s}_2 e^{-i\vec{k}_2 \vec{r}_1} | i \rangle \qquad (2.2b)$$

where $\vec{k}_2 = \omega \vec{v}_2$ and $\vec{s}_2$ are the momentum and polarization vectors of the outgoing photon and $\mu_f$ is the polarization of the electron in the final state and

$$G(\vec{r}_2,\vec{r}_1;\Omega) = S_n \frac{|n\rangle\langle n|}{E_n - \Omega}$$

is the coulombian Green function of the Dirac equation.

The ground state Dirac spinor is

$$u^+_{\mu_f}(\vec{r}_2) = \sqrt{\frac{\lambda^3}{\pi} \frac{\gamma+1}{\Gamma(2\gamma+1)}} (2\lambda r)^{\gamma-1} e^{-\lambda r} \left(1 - \frac{i\alpha Z}{1+\gamma} \frac{\vec{\alpha}\,\vec{r}}{r}\right) \chi^+_\mu \qquad (2.3)$$

where

$$\chi_\mu^+ = \begin{cases} (1\,0\,0\,0) & \text{if } \mu = \frac{1}{2} \\ (0\,1\,0\,0) & \text{if } \mu = -\frac{1}{2} \end{cases}$$

Also, denoting $E_0 = \gamma m$ the relativistic ground state electron energy,

$$\Omega_1 = E_0 + \omega + i\varepsilon\,;\ \Omega_2 = E_0 - \omega - i\varepsilon \tag{2.4}$$

We observe that $\mathcal{M}_{\mu_f \mu_i}(\Omega_2)$ is obtained from $\mathcal{M}_{\mu_f \mu_i}(\Omega_1)$ by replacing $\Omega_1$ with $\Omega_2$ and interchanging $\vec{k}_2$ and $-\vec{k}_1$ as well as $\vec{s}_1$ and $\vec{s}_2$, and both of them have a resonant structure giving rise to physical singularities, as discussed in ref. [7]. We point out that in the specific case of Rayleigh scattering the intermediate states belonging to continuum spectrum for which $E_n \cong E_0 + \omega$ or $E_n \cong E_0 - \omega$ have a large contribution to the matrix elements $\mathcal{M}(\Omega_1)$ and $\mathcal{M}(\Omega_2)$ respectively. In the case of low energies $\omega$, these resonant intermediate states have not so high $E_n$ values and a nonrelativistic KHW approach gives fairly good results. A high energy regime leads to high values for $E_n$ having large contributions to the matrix elements $\mathcal{M}(\Omega)$ and such intermediate states must be considered relativistic, *i.e.* as solutions of Dirac equation with Coulomb field. This means that the sum over intermediate states of positive and negative frequencies has to be expressed in terms of relativistic coulombian Green function $G(\vec{r}_2, \vec{r}_1; \Omega)$ which in the Sommerfeld Maue approach is given by the relationship [8,9]

$$G(\vec{r}_2, \vec{r}_1; \Omega) = \frac{1}{2m}\left(\vec{\alpha}\vec{P}_2 + \beta m + \frac{\alpha Z}{r_2} + \Omega\right)\left[I - \frac{1}{2\Omega}\vec{\alpha}\left(\vec{P}_2 + \vec{P}_1\right)\right]G_0(\vec{r}_2, \vec{r}_1; \Omega) \tag{2.5}$$

where $G_0(\vec{r}_2, \vec{r}_1; \Omega)$ satisfies the Schrodinger-type equation [1]

$$\left(H_0^{Schr}(\vec{r}_2) + \frac{X^2}{2m}\right)G_0(\vec{r}_2, \vec{r}_1; \Omega) = \delta(\vec{r}_2 - \vec{r}_1) \tag{2.6}$$

with

$$X^2 = m^2 - \Omega^2 \tag{2.7}$$

$$H_0^{Schr}(\vec{r}_2) = -\frac{1}{2m}\Delta_2 - \frac{\alpha Z \Omega}{m}\frac{1}{r_2} \tag{2.8}$$

so that $G_0(\vec{r}_2, \vec{r}_1; \Omega)$ has the same expression as the nonrelativistic coulombian Green function but with modified parameters according to relativistic kinematics.

From eqs. (2.2) and (2.5) we get

$$\mathcal{M}_{\mu_f \mu_i}(\Omega_1) = -\frac{1}{2}\langle E_0 \mu_f | \vec{\alpha}\vec{s}_2\, e^{-i\vec{k}_2 \vec{r}_2}\left(\vec{\alpha}\vec{P}_2 + \beta m + \frac{\alpha Z}{r_2} + \Omega\right)\left[I - \frac{1}{2\Omega_1}\vec{\alpha}\left(\vec{P}_1 + \vec{P}_2\right)\right]G_0(\vec{r}_2, \vec{r}_1; \Omega_1)\vec{\alpha}\vec{s}_1\, e^{i\vec{k}_1 \vec{r}_1} | E_0 \mu_i\rangle \tag{2.9}$$

Taking into account that

$$\vec{\alpha}\vec{s}_2\left(\vec{\alpha}\vec{P}_2 + \beta m + \frac{\alpha Z}{r_2} + \Omega\right) = \left(-\vec{\alpha}\vec{P}_2 - \beta m + \frac{\alpha Z}{r_2} + \Omega\right)\vec{\alpha}\vec{s}_2 + 2\vec{s}_2\vec{P}_2$$

and that $u_{\mu_f}^+(\vec{r}_2)$ is a solution of Dirac equation

---

[1] By definition, the function $G_0$ in eq. (2.7) differs from Hostler's function $G_0^{Host}$ by a factor $-2m$: $G_0 = -2m G_0^{Host}$.

$$u^+_{\mu_f}(\vec{r}_2)\left(\vec{\alpha}\vec{P}_2 + m\beta - \frac{\alpha Z}{r_2}\right) = E_0 u^+_{\mu_f}(\vec{r}_2)$$

we get

$$\mathcal{M}_{\mu_f \mu_i}(\Omega_1, \theta) = -\frac{1}{2}\iint\limits_{\mathbb{R}^3 \mathbb{R}^3} d^3r_1 d^3r_2\, u^+_{\mu_f}(\vec{r}_2) e^{-i\vec{k}_2 \vec{r}_2}\left[\left(\Omega_1 - E_0 - \vec{\alpha}\vec{k}_2\right)\vec{\alpha}\vec{s}_2 + 2(\vec{s}_2 \vec{P}_2)\right]$$

$$\times\left[I - \frac{1}{2\Omega_1}\vec{\alpha}\left(\vec{P}_1 + \vec{P}_2\right)\right] G_0(\vec{r}_2, \vec{r}_1; \Omega_1)(\vec{\alpha}\vec{s}_1) e^{i\vec{k}_1 \vec{r}_1} u_{\mu_i}(\vec{r}_1) \quad (2.10)$$

In the same way we get:

$$\mathcal{M}_{\mu_f \mu_i}(\Omega_2, \theta) = -\frac{1}{2}\iint\limits_{\mathbb{R}^3 \mathbb{R}^3} d^3r_1 d^3r_2\, u^+_{\mu_f}(\vec{r}_2) e^{i\vec{k}_1 \vec{r}_2}\left[\left(\Omega_2 - E_0 - \vec{\alpha}\vec{k}_1\right)\vec{\alpha}\vec{s}_1 + 2(\vec{s}_1 \vec{P}_2)\right]$$

$$\times\left[I - \frac{1}{2\Omega_2}\vec{\alpha}\left(\vec{P}_1 + \vec{P}_2\right)\right] G_0(\vec{r}_2, \vec{r}_1; \Omega_2)(\vec{\alpha}\vec{s}_2) e^{-i\vec{k}_2 \vec{r}_1} u_{\mu_i}(\vec{r}_1) \quad (2.11)$$

so that the eq. (2.1) leads to the relationship:

$$\mathcal{M}^R_{\mu_f \mu_i} = -s_{1j}s_{2k}\left[\vartheta_{jk} + \Pi_{jk}(\Omega_1) + \Pi_{jk}(\Omega_2)\right] \quad (2.12)$$

where

$$s_{1j}s_{2k}\vartheta_{jk} = \frac{1}{2}\iint\limits_{\mathbb{R}^3 \mathbb{R}^3} d^3r_1 d^3r_2\, u^+_{\mu_f}(\vec{r}_2)\left\{\vec{\alpha}\vec{s}_2\left(\Omega_1 - E_0 + \vec{\alpha}\vec{k}_2\right)\left[I - \frac{1}{2\Omega_1}\vec{\alpha}\left(\vec{P}_1 + \vec{P}_2\right)\right]G_0(\vec{r}_2, \vec{r}_1; \Omega_1)\vec{\alpha}\vec{s}_1 e^{i(\vec{k}_1 \vec{r}_1 - \vec{k}_2 \vec{r}_2)}\right.$$

$$\left. + \vec{\alpha}\vec{s}_1\left(\Omega_2 - E_0 + \vec{\alpha}\vec{k}_1\right)\left[I - \frac{1}{2\Omega_2}\vec{\alpha}\left(\vec{P}_1 + \vec{P}_2\right)\right]G_0(\vec{r}_2, \vec{r}_1; \Omega_2)\vec{\alpha}\vec{s}_2 e^{i(\vec{k}_1 \vec{r}_2 - \vec{k}_2 \vec{r}_1)}\right\} u_{\mu_i}(\vec{r}_1),$$

(2.13)

$$s_{1j}s_{2k}\Pi_{jk}(\Omega_1) = \iint\limits_{\mathbb{R}^3 \mathbb{R}^3} d^3r_1 d^3r_2\, u^+_{\mu_f}(\vec{r}_2)\vec{s}_2 \vec{P}_2\left[I - \frac{1}{2\Omega_1}\vec{\alpha}\left(\vec{P}_1 + \vec{P}_2\right)\right]G_0(\vec{r}_2, \vec{r}_1; \Omega_1)\vec{\alpha}\vec{s}_1 e^{i(\vec{k}_1 \vec{r}_1 - \vec{k}_2 \vec{r}_2)} u_{\mu_i}(\vec{r}_1), \quad (2.14)$$

$$s_{1j}s_{2k}\Pi_{jk}(\Omega_2) = \iint\limits_{\mathbb{R}^3 \mathbb{R}^3} d^3r_1 d^3r_2\, u^+_{\mu_f}(\vec{r}_2)\vec{s}_1 \vec{P}_2\left[I - \frac{1}{2\Omega_2}\vec{\alpha}\left(\vec{P}_1 + \vec{P}_2\right)\right]G_0(\vec{r}_2, \vec{r}_1; \Omega_2)\vec{\alpha}\vec{s}_2 e^{i(\vec{k}_1 \vec{r}_2 - \vec{k}_2 \vec{r}_1)} u_{\mu_i}(\vec{r}_1).$$

(2.15)

The eqs. (2.10-2.15) are similar to those that appear in the case of Compton scattering [7] unless the final state Dirac spinor, which in the Rayleigh case corresponds to the ground state[2]. In the case of Rayleigh scattering it was shown [3,6] that for low and intermediate Z values and for photon energies up to several times the photoeffect threshold the nonrelativistic approach including multipoles and retardation gives excellent predictions. That suggests that above the photoeffect threshold, in a large range of energies, the first iteration term to the main term $G_0$ should be significantly smaller than the main term itself. We can expect that this is also true for heavy atoms for photon energies $\omega$ up to twice the photoeffect threshold. Comparing the numerical results obtained when the spurious poles are no longer present with the full relativistic numerical results of Kissel *et al* [2] we can check the validity of this presumption. In the following we neglect the first iterative term $G_0$, as well as the "small" components of both ground state Dirac spinors. We discuss in more detail this last approximation in the Appendix A. We want to state that neglecting small components is a consistent approach in the nonrelativistic region.

---

[2] The different sign in the right side of the eq. (2.12) is due to the different sign convention for the function $G_0(\vec{r}_2, \vec{r}_1; \Omega_2)$. The coefficient ½ in the right side of eqs. (2.14) and (2.15) of the reference [7], generated by an edit error, should be dropped out.

Just as in the case of Compton scattering [7] it is possible to write
$$(\Omega - E_0)G_0(\vec{r}_2, \vec{r}_1; \Omega) = -\delta(\vec{r}_2 - \vec{r}_1) + (H_0^{Schr}(\vec{r}_2, \Omega) - E_0)G_0(\vec{r}_2, \vec{r}_1; \Omega)$$
and taking into account that $(\vec{\alpha}\vec{s}_2)(\vec{\alpha}\vec{s}_1) + (\vec{\alpha}\vec{s}_1)(\vec{\alpha}\vec{s}_2) = 2(\vec{s}_1\vec{s}_2)I$, from eq. (2.13) we get

$$-s_{1j}s_{2k}\vartheta_{jk} = \vartheta(\vec{s}_1\vec{s}_2) + s_{1j}s_{2k}\vartheta_{jk} \tag{2.16}$$

where
$$\vartheta = \int_{\mathbb{R}^3} d^3r\, u_{\mu_i}^+(\vec{r}) e^{i\vec{r}\vec{\Delta}} u_{\mu_i}(\vec{r}) = \langle i|e^{i\vec{r}\vec{\Delta}}|i\rangle \tag{2.17}$$

is the relativistic form factor (it is obvious that $\vartheta$ is nonvanishing only if $\mu_f = \mu_i$) with $\vec{\Delta} = \vec{k}_2 - \vec{k}_1$, $\Delta = 2\omega\sin\left(\dfrac{\theta}{2}\right)$ and

$$s_{1j}s_{2k}\vartheta_{jk} = -\frac{1}{2}\iint_{\mathbb{R}^3\mathbb{R}^3} d^3r_1 d^3r_2\, u_{\mu_f}^+(\vec{r}_2)(H_0^{Schr}(\vec{r}_2) - E_0)\Big[G_0(\vec{r}_2,\vec{r}_1;\Omega_1)(\vec{\alpha}\vec{s}_2)(\vec{\alpha}\vec{s}_1) e^{i(\vec{k}_1\vec{r}_1 - \vec{k}_2\vec{r}_2)}$$
$$+ (\vec{\alpha}\vec{s}_1)(\vec{\alpha}\vec{s}_2) G_0(\vec{r}_2,\vec{r}_1;\Omega_2) e^{i(\vec{k}_2\vec{r}_1 - \vec{k}_1\vec{r}_2)}\Big] u_{\mu_i}(\vec{r}_1)$$

where we have omitted the terms $(\vec{\alpha}\vec{k}_1)$ and $(\vec{\alpha}\vec{k}_2)$ because a product of an odd number of Dirac matrices $\vec{\alpha}$ gives no contribution to the integral (in the nonrelativistic approach there are no small components).

For the case of elastic scattering the momentum transfer modulus is $\Delta = 2\omega\sin\left(\dfrac{\theta}{2}\right)$.

We can prove that in the nonrelativistic limit the function $s_{1j}s_{2k}\vartheta_{jk}$ is vanishing. Indeed, in the nonrelativistic limit we have to replace $u_{\mu_f}^+(\vec{r})$ by the solution $u_{NR}^*(\vec{r})\chi_{\mu_f}^+$ of the Schrodinger equation with Coulomb field and it is obvious that
$$(H_0^{Schr}(\vec{r}_2) - E_0)u_{\mu_f}^+(\vec{r}_2) \approx \chi_{\mu_f}^+ (H_0^{Schr}(\vec{r}_2) - E_0)u_{NR}^*(\vec{r}_2) = 0,$$
because if $\omega \leq \alpha Z m$ the neglected term $\dfrac{\alpha^2 Z^2}{r}$ is beyond the first iteration to the main term of the Green function. Thus, in the nonrelativistic region, $s_{1j}s_{2k}\vartheta_{jk} = 0$ and the eq. (2.16) becomes

$$-s_{1j}s_{2k}\vartheta_{jk} = \vartheta(\vec{s}_1\vec{s}_2)$$

Also, because in the nonrelativistic limit we have to replace $u_{\mu_f}(\vec{r})$ with the solution $u_{NR}(\vec{r})$ of the Schrodinger equation with Coulomb field and $\vec{\alpha}\vec{s}$ by $\frac{1}{m}\vec{s}\vec{P}$ [10], from eqs. (2.12) and (2.17) we obtain

$$\mathcal{M}^R_{\mu_f \mu_i} = \left(\vec{s}_1 \vec{s}_2\right)\vartheta - s_{1j}s_{2k}\left[\Pi_{jk}(\Omega_1) + \Pi_{jk}(\Omega_2)\right] \tag{2.18}$$

where the nonrelativistic form factor is (see Appendix A):

$$\vartheta_{NR} = \frac{1}{\left(1 + \frac{\alpha^2 Z^2}{4}\frac{\omega^2}{\omega_{th}^2}\sin^2\frac{\theta}{2}\right)^2} \tag{2.19}$$

which involves the relativistic threshold energy $\omega_{th} = (1-\gamma)m$ and

$$s_{1j}s_{2k}\Pi_{jk}(\Omega_1) = \frac{1}{m}\iint_{\mathbb{R}^3 \mathbb{R}^3} d^3 r_1\, d^3 r_2\, u^*_{NR}(\vec{r}_2)e^{i(\vec{k}_1\vec{r}_1 - \vec{k}_2\vec{r}_2)}\left(\vec{s}_2 \vec{P}_2\right)\left(\vec{s}_1 \vec{P}_1\right)G_0(\vec{r}_2,\vec{r}_1;\Omega_1)u_{NR}(\vec{r}_1) \tag{2.20}$$

$$s_{1j}s_{2k}\Pi_{jk}(\Omega_2) = \frac{1}{m}\iint_{\mathbb{R}^3 \mathbb{R}^3} d^3 r_1\, d^3 r_2\, u^*_{NR}(\vec{r}_2)e^{i(\vec{k}_1\vec{r}_2 - \vec{k}_2\vec{r}_1)}\left(\vec{s}_1 \vec{P}_2\right)\left(\vec{s}_2 \vec{P}_1\right)G_0(\vec{r}_2,\vec{r}_1;\Omega_2)u_{NR}(\vec{r}_1) \tag{2.21}$$

The Green functions $G_0(\vec{r}_2,\vec{r}_1;\Omega_1)$ and $G_0(\vec{r}_2,\vec{r}_1;\Omega_2)$ which are present in the above relationships are given by the eq. (2.7) with

$$X^2(\Omega_1) = m^2 - \Omega_1^2 = -\omega^2 - 2E_0\omega + \alpha^2 Z^2 m^2,\ X^2(\Omega_2) = m^2 - \Omega_2^2 = -\omega^2 + 2E_0\omega + \alpha^2 Z^2 m^2 \tag{2.22}$$

so that $X^2(\Omega_1) + \omega^2 = -2E_0\omega + \alpha^2 Z^2 m^2$ and $X^2(\Omega_2) + \omega^2 = 2E_0\omega + \alpha^2 Z^2 m^2$ are linear and not quadratic functions in the photon energy as the KHM matrix element presents. We notice that, as in the case of Compton scattering [7], the term $\omega^2$ introduced in the denominators (1.10) of Rayleigh amplitudes by nonrelativistic and retardation corrections is canceled by a similar term introduced by relativistic kinematics in the expressions of $X^2(\Omega_1)$ and $X^2(\Omega_2)$. It is obvious that first we have to perform the integrals involved in the amplitudes (2.20) and (2.21) (i.e. the main terms), then to combine the multipoles, retardation and relativistic kinematics terms and only after that the remaining terms in $\omega/m$ and $\alpha^2 Z^2$ due to relativistic kinematics should be neglected.

The calculation of $\Pi_{jk}(\Omega)$ may be performed in momentum space, as in ref. [1]:

$$\Pi_{jk}(\Omega_1) = \frac{1}{m}\iint_{\mathbb{R}^3 \mathbb{R}^3} d^3 p_1 d^3 p_2\, p_{1j}p_{2k}u(\vec{p}_2 - \vec{k}_2)G_0(\vec{p}_2,\vec{p}_1;\Omega_1)u(\vec{p}_1 - \vec{k}_1), \tag{2.23}$$

$$\Pi_{jk}(\Omega_2) = \frac{1}{m}\iint_{\mathbb{R}^3 \mathbb{R}^3} d^3 p_1 d^3 p_2\, p_{1k}p_{2j}u(\vec{p}_2 + \vec{k}_1)G_0(\vec{p}_2,\vec{p}_1;\Omega_2)u(\vec{p}_1 + \vec{k}_2) \tag{2.24}$$

where

$$G_0(\vec{p}_1,\vec{p}_2;\Omega) = \frac{m}{2\pi^2}X^3\left(\frac{ie^{i\pi\tau}}{2\sin\pi\tau}\right)\int_1^{(0+)} d\rho\, \rho^\tau\, \frac{d}{d\rho}\left\{\frac{1-\rho^2}{\rho}\frac{1}{\left[X^2(\vec{p}_1-\vec{p}_2)^2 + (p_1^2 + X^2)(p_2^2 + X^2)\frac{(1-\rho)^2}{4\rho}\right]^2}\right\} \tag{2.25}$$

is the Schrodinger coulombian Green function in momentum representation [5], but with the modified parameters in accordance with the relativistic kinematics (eq. 2.7) $X^2 = m^2 - \Omega^2$, $\text{Re}\, X > 0$ and

$$\tau = \frac{\alpha Z \Omega}{X} \tag{2.26}$$

The function $u(\vec{p}-\vec{k})$ is the ground state eigenfunction in momentum representation:

$$u(\vec{p}-\vec{k}) = \frac{2^{3/2}\lambda^{5/2}}{\pi}\frac{1}{\left((\vec{p}-\vec{k})^2+\lambda^2\right)^2} \tag{2.27}$$

Following the method of calculation of Gavrila and Costescu [1] we get the result (eq. [30] in ref. [1]):

$$s_{1i}s_{2j}\Pi_{ij}(\Omega) = (\vec{s}_1\vec{s}_2)P(\Omega)+(\vec{s}_1\vec{v}_2)(\vec{s}_2\vec{v}_1)Q(\Omega) \tag{2.28}$$

where

$$P(\Omega,\theta) = 128\frac{\lambda^5 X^3}{d^4(\Omega)}\frac{F_1(2-\tau;2,2,3-\tau;x_1,x_2)}{2-\tau} \tag{2.29}$$

$$Q(\Omega,\theta) = \frac{2048\lambda^5 X^5 \omega^2}{d^6(\Omega)}\frac{F_1(3-\tau;3,3,4-\tau;x_1,x_2)}{3-\tau} \tag{2.30}$$

with

$$d(\Omega_1) = 2[\alpha^2 Z^2 m^2 - E_0\omega + \alpha Z m X(\Omega_1)]$$
$$d(\Omega_2) = 2[\alpha^2 Z^2 m^2 + E_0\omega + \alpha Z m X(\Omega_2)] \tag{2.31}$$

From eqs. (2.31) it follows that above the photoeffect threshold $|d(\Omega_1)| = 2m\omega$ and above the pair production threshold $|d(\Omega_2)| = 2m\omega$, so that there are no spurious singularities in the expressions of the amplitudes.

The Appell's functions variables are given by the relationships:

$$x_1 x_2 = p = \left[\frac{d^*(\Omega)}{d(\Omega)}\right]^2 \equiv \xi^2(\Omega) \tag{2.32}$$

$$x_1 + x_2 = s = 2\left[1-2\frac{X^2}{m^2}\sin^2\frac{\theta}{2}\right]\xi(\Omega)$$

where $d^*(\Omega)$ is obtained from $d(\Omega)$ by changing the sign of the last term:

$$d^*(\Omega_1) = 2[\alpha^2 Z^2 m^2 - E_0\omega - \alpha Z m X(\Omega_1)]$$
$$d^*(\Omega_2) = 2[\alpha^2 Z^2 m^2 + E_0\omega - \alpha Z m X(\Omega_2)] \tag{2.33}$$

It is useful to notice that:

$$\xi(\Omega) = \frac{4m^2\omega^2}{d^2(\Omega)} \tag{2.34}$$

If we denote the pair production threshold energy $\omega_{pp} = (1+\gamma)m$, the expressions (2.7) may be also written:

$$X(\Omega_1) = \begin{cases} (-\omega+\omega_{th})^{\frac{1}{2}}(\omega+\omega_{pp})^{\frac{1}{2}} &, \omega<\omega_{th} \\ -i(\omega-\omega_{th})^{\frac{1}{2}}(\omega+\omega_{pp})^{\frac{1}{2}} &, \omega\geq\omega_{th} \end{cases} ; \quad X(\Omega_2) = \begin{cases} (\omega+\omega_{th})^{\frac{1}{2}}(-\omega+\omega_{pp})^{\frac{1}{2}} &, \omega<\omega_{pp} \\ -i(\omega+\omega_{th})^{\frac{1}{2}}(\omega-\omega_{pp})^{\frac{1}{2}} &, \omega\geq\omega_{pp} \end{cases} \tag{2.35}$$

We notice that from eqs. (2.35) it follows $X^2(\Omega) = -\omega^2 \mp 2(m-\omega_{th})\omega + \alpha^2 Z^2 m^2$ where the upper sign (-) corresponds to the case $\Omega = \Omega_1$ and the lower sign (+) to $\Omega = \Omega_2$.

In the nonrelativistic limit, we have to consider $\omega \ll \omega_{pp}$, so that the relativistic terms in $\frac{\omega}{m}$ or smaller should be dropped out. We get

$$X_{NR}(\Omega_1) = \begin{cases} \alpha Z m \left(1 - \frac{\omega}{\omega_{th}}\right)^{1/2}, & \omega < \omega_{th} \\ -i\alpha Z m \left(\frac{\omega}{\omega_{th}} - 1\right)^{1/2}, & \omega_{th} \le \omega \ll \omega_{pp} \end{cases} ; X_{NR}(\Omega_2) = \alpha Z m \left(1 + \frac{\omega}{\omega_{th}}\right)^{1/2}, \omega \ll \omega_{pp}. \quad (2.36)$$

Because the location of the photoeffect threshold $\omega_{th}$ and of all physical resonances is correctly given only by Dirac equation, we consider the exact relativistic value of the K-shell threshold. This allows us a consistent comparison (especially for high Z values), between the S-matrix nonrelativistic limit predictions and the fully relativistic ones, both above and below threshold. Indeed, for high Z elements, there is a difference of about 10% between the relativistic and nonrelativistic threshold energies that would induce a systematic and important discrepancy in any predictions of the nonrelativistic approach. We consider that the consistency of the nonrelativistic approach is not be affected by the choosing of the right experimental value of the threshold energy and the position of the physical resonances.

We should also neglect $\omega - \omega_{th}$ in comparison with $m$ in the nonrelativistic limit, and use the nonrelativistic expression for $X(\Omega)$ so the other initially changed parameter $\tau_1 = \frac{\alpha Z \Omega_1}{X(\Omega)_1} = \frac{\alpha Z m}{X(\Omega)_1}\left(1 + \frac{\omega - \omega_{th}}{m}\right)$ will become:

$$\tau_1 = \frac{\alpha Z m}{X_{NR}(\Omega_1)} = \begin{cases} \left(1 - \frac{\omega}{\omega_{th}}\right)^{-\frac{1}{2}}, & \omega < \omega_{th} \\ i\left(\frac{\omega}{\omega_{th}} - 1\right)^{-\frac{1}{2}}, & \omega_{th} \le \omega \ll \omega_{pp} \end{cases} ; \tau_2 = \frac{\alpha Z m}{X_{NR}(\Omega_2)} = \left(1 + \frac{\omega}{\omega_{th}}\right)^{-\frac{1}{2}}, \omega \ll \omega_{pp} \quad (2.37)$$

We expect this approximation to be valid for energies of the incoming photon up to $2\omega_{th}$ for high Z targets and up to $\alpha Z m$ for low and intermediate Z values.

We want to point out that taking the relativistic value for the threshold energy $\omega_{th} = (1-\gamma)m$ in the expressions (2.37) we obtain very accurately the positions of the physical resonances (apart their fine structure depending on $j$).

From eqs. (2.18) and (2.28) it follows
$$\mathcal{M}^R_{\mu_f \mu_i} = M(\omega,\theta)(\vec{s}_1 \vec{s}_2) + N(\omega,\theta)(\vec{s}_1 \vec{v}_2)(\vec{s}_2 \vec{v}_1) \quad (2.38)$$
where
$$M(\omega,\theta) = [\vartheta - P(\Omega_1,\theta) - P(\Omega_2,\theta)] \quad (2.39)$$
$$N(\omega,\theta) = -[Q(\Omega_1,\theta) + Q(\Omega_2,\theta)] \quad (2.40)$$

**a) The nonrelativistic limit for $\omega \geq \omega_{th}$.**

From eq. (2.32) we get

$$\xi_{NR}(\Omega_1) = e^{2i\chi_0} \tag{2.41}$$

where

$$\chi_0 = \begin{cases} \arctan\left(\dfrac{\sqrt{\dfrac{\omega}{\omega_{th}}-1}}{1-\dfrac{E_0\omega}{\alpha^2 Z^2 m^2}}\right), & \text{if } \omega \leq \dfrac{\alpha^2 Z^2 m^2}{E_0} \\[2em] \pi - \arctan\left(\dfrac{\sqrt{\dfrac{\omega}{\omega_{th}}-1}}{\dfrac{E_0\omega}{\alpha^2 Z^2 m^2}-1}\right), & \text{if } \omega > \dfrac{\alpha^2 Z^2 m^2}{E_0} \end{cases} \tag{2.42}$$

We point out that the presence of the relativistic ground state energy $E_0$ in the above expressions is needed because we used the relativistic photoelectric threshold and also it gives the correct value of the photon energy $\omega$ for which the denominator vanishes. The accuracy of choosing the exact photoelectric threshold value is particularly important here, as the function $\chi_0$ appears in an exponential.

Using the relationship (2.34) and (2.41) we get

$$d_{NR}^2(\Omega_1) = 4m^2\omega^2 e^{-2i\chi_0}$$

For obtaining $u(\Omega_2)$ we start also from eq. (2.32) and we obtain the expression

$$\xi(\Omega_2) = 1 - \frac{2\alpha Z X(\Omega_2)}{\alpha^2 Z^2 m + (1-\omega_{th}/m)\omega + \alpha Z X(\Omega_2)} \tag{2.43}$$

In the nonrelativistic limit we should take the nonrelativistic expression (2.36) for $X(\Omega_2)$ and neglect the term $\omega_{th}/m$ in the denominator of the above equation, which leads to

$$\xi_{NR}(\Omega_2) = \frac{1+\dfrac{\omega}{\alpha^2 Z^2 m}-\left(1+\dfrac{\omega}{\omega_{th}}\right)^{1/2}}{1+\dfrac{\omega}{\alpha^2 Z^2 m}+\left(1+\dfrac{\omega}{\omega_{th}}\right)^{1/2}} \tag{2.44}$$

Using the relationship (2.34) and (2.44) we get

$$\frac{1}{d_{NR}^2(\Omega_2)} = \frac{\xi_{NR}(\Omega_2)}{4m^2\omega^2}; \quad \omega \geq \omega_{th} \tag{2.45}$$

**b) The nonrelativistic limit for $\omega < \omega_{th}$.**

For photon energies below the photoeffect threshold, all parameters are real, thus all amplitudes will also be real.

The locations of the physical resonances can not be described accurately in a nonrelativistic approach. However, the position of the first resonance is fairly well obtained (within 0.3% for Z=47) because we have considered the exact value of the threshold. Properly, all our relationships are useful only below the first resonance (i. e. for $\omega < \tfrac{3}{4}\omega_{th}$).

From the eq. (2.31) it follows:

$$d^2(\Omega) = 4m^2\left[\left(\frac{E_0}{m}\right)^2\omega^2 - \alpha^2 Z^2\omega^2 + 4\alpha^2 Z^2 m\left(\alpha^2 Z^2 m/2 \mp \frac{E_0}{m}\omega\right) + 2\alpha Z(\alpha^2 Z^2 m \mp \frac{E_0}{m}\omega)X(\Omega)\right]$$

In the nonrelativistic limit we have to replace $X(\Omega)$ according to the eq. (2.36) and neglect the term $\alpha^2 Z^2\omega^2$ which compared with the first term is of the order $\omega_{th}/m$, obtaining

$$d^2_{NR}(\Omega) = 4\left[E_0^2\omega^2 + 4\alpha^2 Z^2 m^2(\alpha^2 Z^2 m^2/2 \mp E_0\omega) + 2\alpha^2 Z^2 m^2(\alpha^2 Z^2 m^2 \mp E_0\omega)(1 \mp \omega/\omega_{th})^{1/2}\right] \quad (2.46)$$

We emphasize that especially for high Z elements, it is important to consider the relativistic coulombian values for the ground state energy and photoeffect threshold.

The sum and the products of the variables of the Appell functions, given by eqs. (2.32) will be calculated from these denominators through the function $\xi(\Omega)$ given by eq. (2.34):

$$\xi(\Omega_1) = \frac{\left(\dfrac{\omega}{\omega_{th}}\right)^2}{\left[\left(1+\sqrt{1-\dfrac{\omega}{\omega_{th}}}\right)\left(1+\dfrac{E_0}{m}\right) - \dfrac{E_0}{m}\dfrac{\omega}{\omega_{th}}\right]^2 + \alpha^2 Z^2\dfrac{\omega}{\omega_{th}}} \quad (2.47)$$

In accordance with the symmetry properties of the process we may obtain $\xi(\Omega_2)$ from (2.47) by changing the sign in the front of the energy $\omega$ i.e. $\omega \to -\omega$.

**III. The total cross section of the photoelectric effect**

According to the optical theorem the total photoeffect cross section is related to the imaginary part of Rayleigh amplitude for forward scattering:

$$\sigma_{ph} = \frac{4\pi}{\alpha}\frac{m}{\omega}r_0^2\left|\mathrm{Im}\,\mathcal{M}^R_{\mu_f\mu_i}(\omega)\right|_{\theta=0} = \frac{4\pi}{\alpha}\frac{m}{\omega}r_0^2\left|\mathrm{Im}\,M(\omega)\right|_{\theta=0} \quad (3.1)$$

From eq. (2.39) we get

$$\sigma_{ph} = \frac{4\pi}{\alpha}\frac{m}{\omega}r_0^2\left|\mathrm{Im}\,P(\Omega_1)\right|_{\theta=0} \quad (3.2)$$

We notice that the imaginary part of the amplitudes $P(\Omega_1)$ and $Q(\Omega_1)$ may be expressed in terms of elementary functions for any scattering angle $\theta$. Indeed, there is the identity:

$$\mathrm{Im}\frac{X(\Omega_1)}{d^{2b}(\Omega_1)}\frac{F_1(b-\tau;b,b;b+1-\tau;x,y)}{b-\tau}$$
$$= (-1)^{b+1}\frac{\pi}{(2b-1)!}|X(\Omega_1)|\frac{|\tau|(1+|\tau|^2)..[(b-1)^2+|\tau|^2]}{|d(\Omega_1)|^{2b}}\frac{(-\xi)^\tau}{e^{\pi|\tau|}-e^{-\pi|\tau|}}{}_2F_1(b-\tau;b+\tau;b+1/2;v) \quad (3.3)$$

with

$$v(\theta) = -\frac{1}{2}\left(1-\frac{s(\theta)}{2\xi}\right) = \frac{|X(\Omega_1)|^2}{m^2}\sin^2\frac{\theta}{2} \quad \text{and} \quad (-\xi)^\tau = e^{-|\tau|\chi_R(\omega)}$$

where the exact expression of $\chi_R(\omega)$ is given by:

$$\chi_R(\omega) = \begin{cases} -\pi + 2\arctan\left(\dfrac{\alpha Z |X_1|}{\alpha^2 Z^2 m - E_0 \omega / m}\right), & \text{if} \quad \omega \leq \alpha^2 Z^2 m^2 / E_0 \\ \pi - 2\arctan\left(\dfrac{\alpha Z |X_1|}{E_0 \omega / m - \alpha^2 Z^2 m}\right), & \text{if} \quad \omega > \alpha^2 Z^2 m^2 / E_0 \end{cases} \quad (3.4)$$

In the case of forward scattering, $s = 2\xi$ and $v = 0$ so that the Gauss function in the expression (3.3) equals 1.

$$\operatorname{Im} \frac{X}{d^4(\Omega)} \frac{F_1(2-\tau; 2, 2; 3-\tau; x, y)}{2-\tau} = -\frac{\pi \alpha Z}{96}\left(1+|\tau_1|^2\right)\frac{|\Omega|}{m^4 \omega^4}\frac{e^{-|\tau_1|(\pi+\chi_R(\omega))}}{1-e^{-2\pi|\tau_1|}} \quad (3.5)$$

Introducing eq. (2.34) in (2.29) we get

$$\operatorname{Im} P(\Omega_1)\big|_{\theta=0} = -\frac{4\pi}{3}(\alpha Z)^6 m \frac{\Omega_1 |X_1|^2}{\omega^4}\left(1+|\tau_1|^2\right)\frac{e^{-|\tau_1|(\pi+\chi_R(\omega))}}{1-e^{-2\pi|\tau_1|}}$$

$$\sigma_{ph} = \frac{16}{3}\pi^2 r_0^2 m^2 \alpha^5 Z^6 \frac{(E_0+\omega)|X_1|^2}{\omega^5}\left(1+|\tau_1|^2\right)\frac{e^{-|\tau_1|(\pi+\chi_R(\omega))}}{1-e^{-2\pi|\tau_1|}}$$
(3.6)

We point out that the exponential $e^{-|\tau|\chi(\Omega_1)}$ is just the exponential that appears in the full relativistic calculation of the Rayleigh amplitude when the ground state Dirac spinors are exactly taken into account [11]. This expression is involved both in the calculation of the relativistic formulae for the photoeffect and pair production (with the final electron created in the K-shell) cross sections. Obviously, for the last case, as a consequence of time reversal invariance of the theory, it is necessary to change everywhere the sign in front of the energy $\omega$.

In the nonrelativistic limit we have to take in the expression (3.4) the expression (2.36) of $X_1$, obtaining the nonrelativistic exponent:

$$\chi(\omega) = \begin{cases} -\pi + 2\arctan\left(\dfrac{\sqrt{\omega/\omega_{th}-1}}{1-\dfrac{E_0\omega}{\alpha^2 Z^2 m^2}}\right), & \text{if} \quad \omega \leq \dfrac{\alpha^2 Z^2 m^2}{E_0} \\ \pi - 2\arctan\left(\dfrac{\sqrt{\omega/\omega_{th}-1}}{\dfrac{E_0\omega}{\alpha^2 Z^2 m^2}-1}\right), & \text{if} \quad \omega > \dfrac{\alpha^2 Z^2 m^2}{E_0} \end{cases} \quad (3.7)$$

Also, in the nonrelativistic limit we take the nonrelativistic values of $|X_1|^2$ and $|\tau_1|^2$ given by eqs. (2.22) and (2.37) and neglect the term in $\omega^2$.

We obtain the following expression of the nonrelativistic cross section for K-shell bound electrons photoeffect with multipoles and retardation included:

$$\sigma_{ph} = \frac{32}{3}\pi^2 r_0^2 E_0 m^2 \alpha^5 Z^6 \frac{E_0+\omega}{\omega^4}\frac{e^{-|\tau_1|(\pi+\chi_R(\omega))}}{1-e^{-2\pi|\tau_1|}} \quad (3.8)$$

The predictions given by formula (3.8), presented in the next chapter, show a very good agreement with full relativistic numerical calculations of Kissell *et. al.* [2], Mayers and Brown [12], Scofield [13], Hultberg *et. al.* [14], for any atomic number *Z* and photon energies at least up to twice the photoeffect threshold. We notice that the formula (3.8) gives significantly better predictions than the relativistic Sauter formula [15], due to the presence of the exponential $e^{-|\tau|\chi(\Omega_1)}$.

We consider that our formula (3.8), which obviously has no spurious singularities is more accurate than Fischer's formula for the nonrelativistic photoeffect from K-shell bound electrons and it also works at larger energies and Z values.

### IV. Numerical results and conclusions

In the tables I-VIII we present our angular distribution predictions for Rayleigh scattering by a K-shell electron of the elements with Z=30, Z=47, Z=82 and Z=92, for several photon energies above and below the photoeffect threshold $\omega_{th}$, for the amplitudes $A_\perp = M$, $A_{II} = M\cos\theta - N\sin^2\theta$, where $M$ and $N$ are given by the eqs. (2.39, 2.40). Figures 1 and 2 show a very good agreement with the full relativistic calculation of Kissel *et. al.* [2] for Z=47 both above and below the threshold ([3], figure 3). We expect the same good agreement for higher Z elements, and we present in figure 3 our angular distribution for Pb.

Also, for forward scattering our result are in a very good agreement with other result presented in the literature, as shown in tables IX and X.

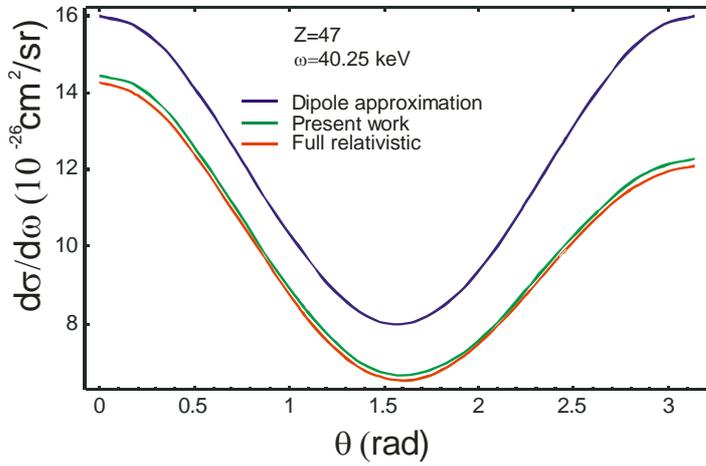

FIG. 1. Angular distribution of the Rayleigh scattering cross section for a K-shell electron of Ag at 40.25 keV; present work (green), the full relativistic results (red) and dipole approximation (blue).

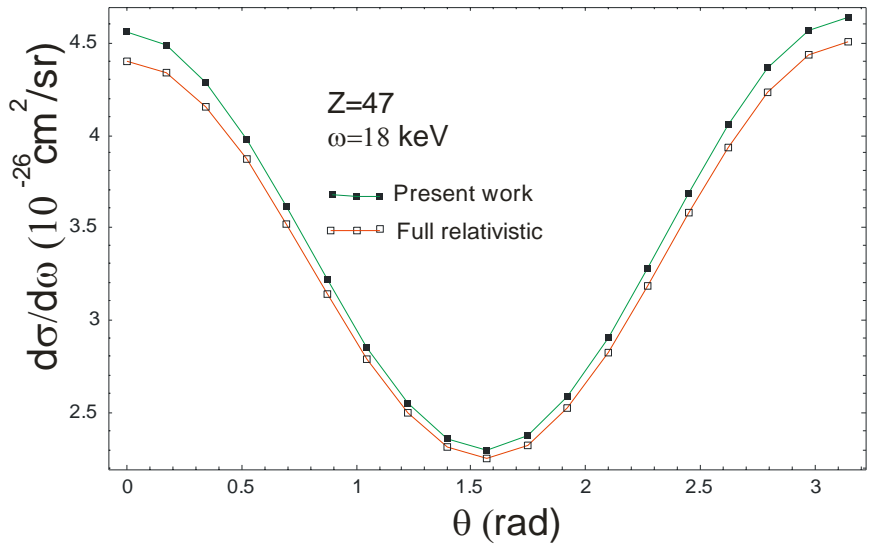

FIG. 2. Angular distribution of the Rayleigh scattering cross section for a K-shell electron of Ag at 18.0 keV; present work (green, solid squares) and full relativistic results (red, squares).

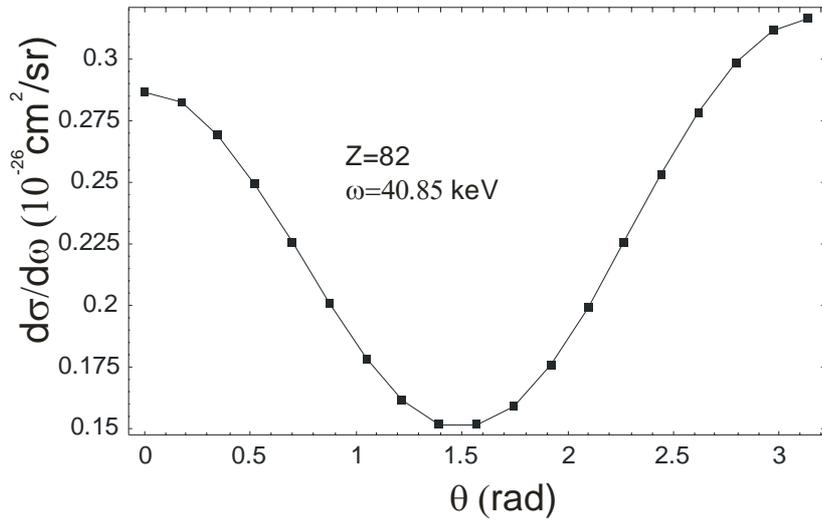

FIG. 3. Angular distribution of the Rayleigh scattering cross section for a K-shell electron of Pb at 40.85 keV, in the nonrelativistic limit.

TABLE I. The angular distribution of Rayleigh cross section (CS) and amplitudes for a K-shell electron of Zn at various photon energies above the photoeffect threshold.

| θ (degr) | ℏω (keV) | CS($10^{-26}$ cm$^2$/sr) | Re $A_\perp$ | Im $A_\perp$ | Re $A_\parallel$ | Im $A_\parallel$ |
|---|---|---|---|---|---|---|
| 0   | 17.400 | 14.7812 | 1.18449  | -0.677229 | 1.18449    | -0.677229 |
| 30  | 17.400 | 12.8424 | 1.18138  | -0.67496  | 1.0232     | -0.580307 |
| 60  | 17.400 | 9.01563 | 1.17296  | -0.668809 | 0.586719   | -0.321873 |
| 90  | 17.400 | 7.08982 | 1.16161  | -0.66052  | 0.0002467  | 0.0164467 |
| 120 | 17.400 | 8.7108  | 1.15042  | -0.652361 | -0.575073  | 0.338323  |
| 150 | 17.400 | 11.9867 | 1.14233  | -0.646469 | -0.989251  | 0.56386   |
| 180 | 17.400 | 13.6038 | 1.13939  | -0.644329 | -1.13939   | 0.644329  |
| 0   | 40.250 | 9.95089 | 1.11076  | -0.139684 | 1.11076    | -0.139684 |
| 30  | 40.250 | 8.39112 | 1.0922   | -0.13791  | 0.942478   | -0.116154 |
| 60  | 40.250 | 5.45079 | 1.0438   | -0.133233 | 0.512326   | -0.0572804 |
| 90  | 40.250 | 3.89887 | 0.982673 | -0.127222 | -0.0117487 | 0.0116069 |
| 120 | 40.250 | 4.3695  | 0.92672  | -0.121613 | -0.471487  | 0.0689373 |
| 150 | 40.250 | 5.60112 | 0.888709 | -0.117741 | -0.772201  | 0.104548  |
| 180 | 40.250 | 6.19164 | 0.875373 | -0.11637  | -0.875373  | 0.11637   |
| 0   | 75.100 | 8.70208 | 1.04615  | -0.0396966 | 1.04615    | -0.0396966 |
| 30  | 75.100 | 6.75905 | 0.987566 | -0.0386368 | 0.851349   | -0.0315242 |
| 60  | 75.100 | 3.56178 | 0.850425 | -0.0359566 | 0.415348   | -0.0127873 |
| 90  | 75.100 | 1.97692 | 0.70481  | -0.0327395 | -0.0105366 | 0.00597863 |
| 120 | 75.100 | 1.76844 | 0.593535 | -0.0299502 | -0.303189  | 0.0188775 |
| 150 | 75.100 | 1.94868 | 0.527918 | -0.028136  | -0.459043  | 0.0255466 |
| 180 | 75.100 | 2.0436  | 0.506585 | -0.0275145 | -0.506585  | 0.0275145 |

TABLE II. The angular distribution of Rayleigh cross section (CS) and amplitudes for a K-shell electron of Ag at various photon energies above the photoeffect threshold.

| θ (degr) | ℏω (keV) | CS ($10^{-26}$ cm$^2$/sr) | Re $A_\perp$ | Im $A_\perp$ | Re $A_\parallel$ | Im $A_\parallel$ |
|---|---|---|---|---|---|---|
| 0   | 40.250  | 14.4556 | 1.11163  | -0.764803  | 1.11163    | 0.764803  |
| 30  | 40.250  | 12.4737 | 1.10616  | -0.759535  | 0.960056   | -0.647987 |
| 60  | 40.250  | 8.61211 | 1.09141  | -0.745363  | 0.551631   | -0.343969 |
| 90  | 40.250  | 6.66099 | 1.07173  | -0.726513  | 0.0072757  | 0.0371267 |
| 120 | 40.250  | 8.04518 | 1.0526   | -0.70823   | -0.521284  | 0.381124  |
| 150 | 40.250  | 10.8873 | 1.03891  | -0.695193  | -0.898153  | 0.610861  |
| 180 | 40.250  | 12.2739 | 1.03397  | -0.690493  | -1.03397   | 0.690493  |
| 0   | 75.100  | 10.3487 | 1.117    | -0.236007  | 1.117      | -0.236007 |
| 30  | 75.100  | 8.6018  | 1.0917   | -0.231105  | 0.940736   | -0.191137 |
| 60  | 75.100  | 5.39711 | 1.02678  | -0.218453  | 0.50038    | -0.0843568 |
| 90  | 75.100  | 3.72786 | 0.946992 | -0.202755  | -0.0155658 | 0.0297181 |
| 120 | 75.100  | 4.03894 | 0.876078 | -0.188658  | -0.448553  | 0.114382  |
| 150 | 75.100  | 5.02467 | 0.829053 | -0.179231  | -0.721234  | 0.16142   |
| 180 | 75.100  | 5.49086 | 0.812773 | -0.175953  | -0.812773  | 0.175953  |
| 0   | 122.500 | 9.05434 | 1.06412  | -0.0895896 | 1.06412    | -0.0895896 |
| 30  | 122.500 | 6.97581 | 1.00175  | -0.0860817 | 0.861162   | -0.0682014 |
| 60  | 122.500 | 3.61532 | 0.856684 | -0.0775049 | 0.412621   | -0.0225432 |
| 90  | 122.500 | 1.98902 | 0.704172 | -0.0677516 | -0.0162842 | 0.0175885 |
| 120 | 122.500 | 1.76397 | 0.58878  | -0.0597583 | -0.304037  | 0.0407675 |
| 150 | 122.500 | 1.9195  | 0.521248 | -0.0547842 | -0.454146  | 0.0506233 |
| 180 | 122.500 | 2.00245 | 0.499381 | -0.0531202 | -0.499381  | 0.0531202 |

TABLE III. The angular distribution of Rayleigh cross section (CS) and amplitudes for a K-shell electron of Pb at various photon energies above the photoeffect threshold.

| θ (degr) | ℏω (keV) | CS ($10^{-26}$cm$^2$/sr) | Re $A_\perp$ | Im $A_\perp$ | Re $A_\parallel$ | Im $A_\parallel$ |
|---|---|---|---|---|---|---|
| 0 | 120.544 | 11.4567 | 0.844681 | -0.854081 | 0.844681 | -0.854081 |
| 30 | 120.544 | 9.74163 | 0.838175 | -0.839899 | 0.744502 | -0.701151 |
| 60 | 120.544 | 6.49473 | 0.820578 | -0.802448 | 0.460642 | -0.326381 |
| 90 | 120.544 | 4.82782 | 0.796984 | -0.754202 | 0.0582894 | 0.0932803 |
| 120 | 120.544 | 5.55782 | 0.773941 | -0.709087 | -0.34901 | 0.419984 |
| 150 | 120.544 | 7.21821 | 0.757436 | -0.677913 | -0.64455 | 0.607871 |
| 180 | 120.544 | 8.01473 | 0.751474 | -0.666874 | -0.751474 | 0.666874 |
| 0 | 139.335 | 10.021 | 0.924275 | -0.63862 | 0.924275 | -0.63862 |
| 30 | 139.335 | 8.41441 | 0.909353 | -0.623608 | 0.800699 | -0.512461 |
| 60 | 139.335 | 5.42258 | 0.870248 | -0.584779 | 0.469024 | -0.215957 |
| 90 | 139.335 | 3.85384 | 0.820511 | -0.536462 | 0.0366743 | 0.0915844 |
| 120 | 139.335 | 4.25352 | 0.774645 | -0.492984 | -0.365006 | 0.308384 |
| 150 | 139.335 | 5.36349 | 0.743305 | -0.463882 | -0.637343 | 0.42088 |
| 180 | 139.335 | 5.89234 | 0.732275 | -0.453759 | -0.732275 | 0.453759 |
| 0 | 158.926 | 9.29345 | 0.964699 | -0.489735 | 0.964699 | -0.489735 |
| 30 | 158.926 | 7.69032 | 0.941027 | -0.47474 | 0.824009 | -0.38373 |
| 60 | 158.926 | 4.76644 | 0.880513 | -0.436767 | 0.46218 | -0.144785 |
| 90 | 158.926 | 3.22081 | 0.806628 | -0.391137 | 0.021819 | 0.0848034 |
| 120 | 158.926 | 3.39551 | 0.741449 | -0.351593 | -0.358541 | 0.231074 |
| 150 | 158.926 | 4.15037 | 0.698489 | -0.325919 | -0.601649 | 0.298937 |
| 180 | 158.926 | 4.50963 | 0.683667 | -0.317137 | -0.683667 | 0.317137 |
| 0 | 199.907 | 8.61504 | 0.995307 | -0.307252 | 0.995307 | -0.307252 |
| 30 | 199.907 | 6.88268 | 0.952758 | -0.293355 | 0.829145 | -0.228962 |
| 60 | 199.907 | 3.89251 | 0.849344 | -0.259652 | 0.432751 | -0.0665225 |
| 90 | 199.907 | 2.34643 | 0.732959 | -0.221858 | 0.0060168 | 0.0675892 |
| 120 | 199.907 | 2.23819 | 0.638689 | -0.191376 | -0.317027 | 0.136874 |
| 150 | 199.907 | 2.56207 | 0.580601 | -0.172665 | -0.502397 | 0.161421 |
| 180 | 199.907 | 2.72128 | 0.561276 | -0.166455 | -0.561276 | 0.166455 |

TABLE IV. The angular distribution of Rayleigh cross section and amplitudes for a K-shell electron of U at 158.926 keV photon energy (above the photoeffect threshold).

| θ (deg) | CS ($10^{-26}$cm$^2$/sr) | Re $A_\perp$ | Im $A_\perp$ | Re $A_\parallel$ | Im $A_\parallel$ |
|---|---|---|---|---|---|
| 0 | 9.77864 | 0.758742 | -0.809879 | 0.758742 | -0.809879 |
| 30 | 8.27448 | 0.752646 | -0.792429 | 0.679732 | -0.654095 |
| 60 | 5.45333 | 0.735701 | -0.746847 | 0.441034 | -0.283051 |
| 90 | 3.97366 | 0.712161 | -0.689196 | 0.081348 | 0.110283 |
| 120 | 4.44753 | 0.688515 | -0.636399 | -0.29329 | 0.393995 |
| 150 | 5.6687 | 0.671299 | -0.600554 | -0.566458 | 0.543814 |
| 180 | 6.25662 | 0.665036 | -0.587989 | -0.665036 | 0.587989 |

TABLE V. The angular distribution of Rayleigh cross section and amplitudes for a K-shell electron of Zn at various photon energies below the photoeffect threshold.

| θ (deg) | ℏω (keV) | CS ($10^{-26}$cm$^2$/sr) | Re $A_\perp$ | Re $A_\parallel$ |
|---|---|---|---|---|
| 0   | 1.38 | 0.00152463  | -0.0138572 | -0.0138572 |
| 30  | 1.38 | 0.00133408  | -0.0138669 | -0.0119899 |
| 60  | 1.38 | 0.000954683 | -0.0138932 | -0.00688909 |
| 90  | 1.38 | 0.000770263 | -0.0139291 | 0.000076640 |
| 120 | 1.38 | 0.00097097  | -0.013965  | 0.00703997 |
| 150 | 1.38 | 0.00136183  | -0.0139913 | 0.012136 |
| 180 | 1.38 | 0.00155641  | -0.0140009 | 0.0140009 |
| 0   | 5.41 | 0.687231    | -0.294202  | -0.294202 |
| 30  | 5.41 | 0.60118     | -0.29433   | -0.254566 |
| 60  | 5.41 | 0.429756    | -0.29468   | -0.146346 |
| 90  | 5.41 | 0.345854    | -0.295156  | 0.00132333 |
| 120 | 5.41 | 0.434868    | -0.295631  | 0.148807 |
| 150 | 5.41 | 0.60928     | -0.295978  | 0.256654 |
| 180 | 5.41 | 0.696147    | -0.296105  | 0.296105 |
| 0   | 8.04 | 17.0186     | -1.46405   | -1.46405 |
| 30  | 8.04 | 14.8861     | -1.46424   | -1.26717 |
| 60  | 8.04 | 10.6314     | -1.46477   | -0.729686 |
| 90  | 8.04 | 8.52592     | -1.46548   | 0.00358761 |
| 120 | 8.04 | 10.6832     | -1.46618   | 0.735775 |
| 150 | 8.04 | 14.9541     | -1.46669   | 1.27109 |
| 180 | 8.04 | 17.0844     | -1.46688   | 1.46688 |

TABLE VI. The angular distribution of Rayleigh cross section and amplitudes for a K-shell electron of Ag at various photon energies below the photoeffect threshold.

| θ (deg) | ℏω (keV) | CS ($10^{-26}$cm$^2$/sr) | Re $A_\perp$ | Re $A_\parallel$ |
|---|---|---|---|---|
| 0   | 5.41  | 0.00900703 | -0.033681  | -0.033681 |
| 30  | 5.361 | 0.00787893 | -0.0337361 | -0.0290952 |
| 60  | 5.41  | 0.00565    | -0.0338866 | -0.0165801 |
| 90  | 5.41  | 0.00461504 | -0.0340921 | 0.000484061 |
| 120 | 5.41  | 0.00588722 | -0.0342973 | 0.0175115 |
| 150 | 5.41  | 0.0082726  | -0.0344474 | 0.0299533 |
| 180 | 5.41  | 0.00945168 | -0.0345024 | 0.0345024 |
| 0   | 17.43 | 3.45116    | -0.659291  | 3.45116 |
| 30  | 17.43 | 3.0168     | -0.65972   | 3.0168 |
| 60  | 17.43 | 2.15555    | -0.660884  | 2.15555 |
| 90  | 17.43 | 1.74234    | -0.662458  | 1.74234 |
| 120 | 17.43 | 2.19996    | -0.664014  | 2.19996 |
| 150 | 17.43 | 3.08046    | -0.665142  | 3.08046 |
| 180 | 17.43 | 3.51703    | -0.665552  | 3.51703 |
| 0   | 22.1  | 126.152    | -3.98604   | -3.98604 |
| 30  | 22.1  | 110.324    | -3.98644   | -3.44939 |
| 60  | 22.1  | 78.7639    | -3.98753   | -1.98489 |
| 90  | 22.1  | 63.1698    | -3.98899   | 0.0117565 |
| 120 | 22.1  | 79.1565    | -3.9904    | 2.00396 |
| 150 | 22.1  | 110.76     | -3.9914    | 3.45956 |
| 180 | 22.1  | 126.515    | -3.99177   | 3.99177 |
| 0   | 26.0  | 5.8997     | 0.862004   | 0.862004 |
| 30  | 26.0  | 5.20771    | 0.862643   | 0.75342 |
| 60  | 26.0  | 3.77442    | 0.864415   | 0.451158 |
| 90  | 26.0  | 2.98592    | 0.866895   | 0.0250922 |
| 120 | 26.0  | 3.6881     | 0.869443   | -0.416032 |
| 150 | 26.0  | 5.23778    | 0.871349   | -0.748412 |
| 180 | 26.0  | 6.03809    | 0.872055   | -0.872055 |

TABLE VII. The angular distribution of Rayleigh cross section and amplitudes for a K-shell electron of Pb at various photon energies below the photoeffect threshold.

| θ (deg) | ℏω (keV) | CS ($10^{-26}$ cm$^2$/sr) | Re $A_\perp$ | Re $A_\parallel$ |
|---|---|---|---|---|
| 0 | 17.43 | 0.00603438 | -0.0275683 | -0.0275683 |
| 30 | 17.43 | 0.00525361 | -0.0277013 | -0.0235795 |
| 60 | 17.43 | 0.00377731 | -0.0280641 | -0.012802 |
| 90 | 17.43 | 0.00324847 | -0.0285586 | 0.00163742 |
| 120 | 17.43 | 0.00433567 | -0.0290518 | 0.015752 |
| 150 | 17.43 | 0.00609315 | -0.029412 | 0.0258798 |
| 180 | 17.43 | 0.00693013 | -0.0295437 | 0.0295437 |
| 0 | 59.60 | 3.48813 | -0.662812 | -0.662812 |
| 30 | 59.60 | 3.03765 | -0.663964 | -0.569491 |
| 60 | 59.60 | 2.16582 | -0.66705 | -0.317178 |
| 90 | 59.60 | 1.78994 | -0.671131 | 0.0214254 |
| 120 | 59.60 | 2.30472 | -0.67506 | 0.353328 |
| 150 | 59.60 | 3.21646 | -0.677843 | 0.592231 |
| 180 | 59.60 | 3.65889 | -0.678842 | 0.678842 |
| 0 | 74.96 | 801.95 | -10.05 | -10.05 |
| 30 | 74.96 | 701.165 | -10.0511 | -8.69449 |
| 60 | 74.96 | 500.434 | -10.054 | -4.99738 |
| 90 | 74.96 | 401.581 | -10.0576 | 0.0385509 |
| 120 | 74.96 | 503.422 | -10.0608 | 5.05863 |
| 150 | 74.96 | 704.149 | -10.0629 | 8.72402 |
| 180 | 74.96 | 804.129 | -10.0637 | 10.0637 |
| 0 | 84.26 | 8.5367 | 1.03691 | 1.03691 |
| 30 | 84.26 | 7.58772 | 1.03721 | 0.91406 |
| 60 | 84.26 | 5.54986 | 1.03825 | 0.565702 |
| 90 | 84.26 | 4.30979 | 1.04016 | 0.0606027 |
| 120 | 84.26 | 5.21838 | 1.04261 | -0.476923 |
| 150 | 84.26 | 7.47878 | 1.04471 | -0.8902 |
| 180 | 84.26 | 8.6794 | 1.04554 | -1.04554 |

TABLE VIII. The angular distribution of Rayleigh cross section and amplitudes for a K-shell electron of U at 84.26 keV photon energy (below the photoeffect threshold).

| θ (deg) | ℏω (keV) | CS ($10^{-26}$ cm$^2$/sr) | Re $A_\perp$ | Re $A_\parallel$ |
|---|---|---|---|---|
| 0 | 84.26 | 7.0897 | -0.944949 | -0.944949 |
| 30 | 84.26 | 6.16551 | -0.946355 | -0.810846 |
| 60 | 84.26 | 4.38522 | -0.950087 | -0.449387 |
| 90 | 84.26 | 3.62455 | -0.954931 | 0.0333177 |
| 120 | 84.26 | 4.66363 | -0.959496 | 0.504095 |
| 150 | 84.26 | 6.49128 | -0.962668 | 0.841659 |
| 180 | 84.26 | 7.3753 | -0.963794 | 0.963794 |

TABLE IX. Comparison of the Rayleigh scattering amplitude $A_\perp$ obtained by us in the nonrelativistic limit to the relativistic S-matrix results of Kissel *et. al.* [2] (RMP) and relativistic dipole results of Florescu *et. al.* [10] (RDLWL) for Z=47, θ=0.

| ℏω (keV) | $k_{REL}$ | $A_\perp$ (present work) | $A_\perp$ (RMP ref [2] ) | $A_\perp$ (RDPLW ref [10] ) | ε(%) |
|---|---|---|---|---|---|
| 5.41 | 0.174544 | -0.033681 | -0.0335 | -0.033576 | 0.54 |
| 17.43 | 0.562347 | -0.659291 | -0.649 | -0.65813 | 1.58 |
| 22.10 | 0.713016 | -3.98604 | -3.713 | -3.7967 | 7.35 |
| 26.0 | 0.838842 | 0.862004 | 0.895 | 0.95226 | -3.68 |

TABLE X. Comparison of the Rayleigh scattering amplitude $A_\perp$ obtained by us in the nonrelativistic limit to the relativistic S-matrix results of Kissel *et. al.* [2] (RMP) and relativistic dipole results of Florescu *et. al.* [10] (RDLWL) for Z=82, θ=0.

| ℏω (keV) | $k_{REL}$ | $A_\perp$ (present work) | $A_\perp$ (RMP ref [2] ) | $A_\perp$ (RDPLW ref [10] ) | ε(%) |
|---|---|---|---|---|---|
| 5.41 | 0.0532576 | -0.00255512 | -0.00254 | -0.0025443 | 0.59 |
| 17.43 | 0.171586 | -0.0275683 | -0.027 | -0.027484 | 2.1 |
| 40.85 | 0.402139 | -0.190065 | -0.187 | -0.19075 | 1.91 |
| 59.5 | 0.585735 | -0.657714 |  | -0.65506 | 0.4 |
| 74.96 | 0.737928 | -10.05 | -10.536 | -10.450 | -4.60 |
| 84.26 | 0.82948 | 1.03691 | 1.2155 | 1.4277 | -14.65 |

In figure 4 we present a comparison of the total photoelectric cross section given by the formula 3.8 with other published relativistic data, showing a very good agreement (estimated errors under 2%).

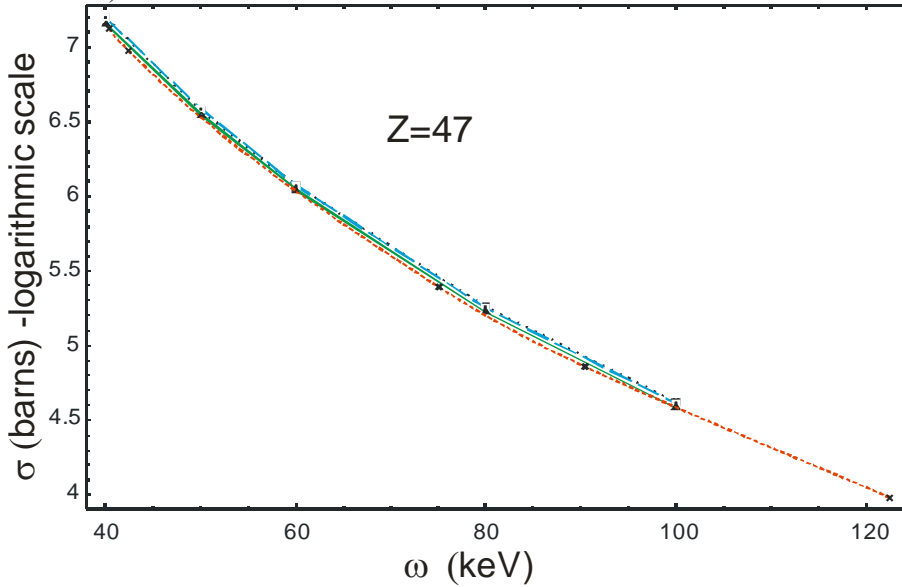

FIG. 4. A comparison of our result for the K-shell photoeffect total cross section (green, continues line) with full relativistic numerical calculations of Kissell *et. al.* [2] (red, small dashed line), Scofield [13] (blue, large dashed line), Hultberg *et. al.* [14] (black, dotted line).

We also present in table XI our predictions for the K-shell photoeffect total cross section for various Z and incoming photon energies.

TABLE XI. K-shell photoeffect total cross section calculated with formula 3.8 for low, intermediate and high Z elements.

| ω (keV) | $\sigma_{ph}$ (barn) | | | | | |
|---|---|---|---|---|---|---|
| | Z=12 | Z=30 | Z=47 | Z=56 | Z=82 | 92 |
| 5.410 | 2539.01 | – | – | – | – | – |
| 7.430 | 985.03 | – | – | – | – | – |
| 22.100 | 31.38 | 1393.49 | – | – | – | – |
| 26.000 | 18.35 | 880.02 | – | – | – | – |
| 37.120 | 5.57 | 315.26 | 1635.23 | – | – | – |
| 40.850 | 4.03 | 238.14 | 1257.10 | – | – | – |
| 59.500 | 1.11 | 77.80 | 442.60 | 828.48 | – | – |
| 74.960 | – | 38.66 | 231.07 | 437.47 | – | – |
| 84.260 | – | 27.05 | 165.87 | 316.16 | – | – |
| 120.544 | – | 8.97 | 59.55 | 116.29 | 456.73 | – |
| 139.335 | – | – | 39.21 | 77.41 | 305.92 | 460.04 |
| 158.926 | – | – | 26.79 | 53.45 | 212.98 | 318.91 |
| 199.907 | – | – | 13.76 | 27.97 | 113.72 | 169.40 |

The good agreement of our calculation with the full relativistic results shows that, for the presented energies regime, the main relativistic kinematics terms are cancelled by retardation and multipoles terms, and the remaining terms have a nonrelativistic origin. Also, the spin effects are small but they settle on the right locations and number of physical resonances. For high Z values, even in the nonrelativistic limit where the nonrelativistic terms are largely dominant, all resonances and threshold energies values must be considered in accordance with their relativistic coulombian values.

**ACKNOWLEDGEMENTS**

This work was partially supported by the Romanian National Research Authority (ANCS) under Grant CEEX PC-D11-PT00-582/2005.
One of the authors (A.C.) is warmly grateful to Professor R.H. Pratt, Dr. Lynn Kissel and Dr. P.M. Bergstrom, for many valuable discussions and useful comments on the subject, while visiting the Department of Physics and Astronomy of the University of Pittsburgh.

**APPENDIX A**

In momentum space eq. (2.9) may be written

$$\mathcal{M}_{\mu_f \mu_i}(\Omega_1, \theta) = \frac{1}{2} \iint_{\mathbb{R}^3 \mathbb{R}^3} d^3 p_1 d^3 p_2\, u^+_{\mu_f}(\vec{p}_2 - \vec{k}_2)\, e^{-i\vec{k}_2 \vec{r}_2} \left[ \left( \Omega - E_0 - \vec{\alpha}\vec{k}_2 \right) \vec{\alpha}\vec{s}_2 \right.$$
$$\left. + 2(\vec{s}_2 \vec{p}_2) \right] \left[ I + \frac{1}{2\Omega_1} \vec{\alpha}\left( \vec{P}_1 + \vec{P}_2 \right) \right] G_0(\vec{p}_2, \vec{p}_1; \Omega_1)(\vec{\alpha}\vec{s}_1)\, e^{i\vec{k}_1 \vec{r}_1}\, u_{\mu_i}(\vec{p}_1 - \vec{k}_1) \quad (A.1)$$

where $u_\mu(\vec{p})$ is the K-shell eigen spinor in momentum space and $G_0(\vec{p}_2, \vec{p}_1; \Omega_1)$ is the relativistic coulombian Green function in momentum space. The Dirac spinor has the form

$$u_\mu(\vec{p}-\vec{k}) = \left[ a(\vec{p}-\vec{k}) + \frac{1}{2} b(\vec{p}-\vec{k}) \frac{\vec{\alpha}\cdot(\vec{p}-\vec{k})}{m} \right] \chi_\mu ;$$

$$\chi^+_{\mu_2} = (1000) \text{ if } \mu=1/2 \text{ and } \chi^+_{\mu_2} = (0100) \text{ if } \mu=-1/2 \quad (A.2)$$

with

$$a(\vec{p}-\vec{k}) = \frac{N\lambda}{\Gamma(2-\gamma)} \int_0^\infty x^{1-\gamma} \frac{d}{dx}\left\{ \frac{1+x}{\left[(\vec{p}-\vec{k})^2 + \lambda^2(1+x)^2\right]^2} \right\} dx \quad (A.3)$$

$$b(\vec{p}-\vec{k}) = \frac{N\lambda}{\Gamma(2-\gamma)} \frac{2}{1+\gamma} \int_0^\infty x^{1-\gamma} \frac{d}{dx}\left\{ \frac{1}{\left[(\vec{p}-\vec{k})^2 + \lambda^2(1+x)^2\right]^2} \right\} dx \quad (A.4)$$

and

$$N = \frac{2^{\gamma+\frac{1}{2}}}{\pi} \lambda^{3/2} \left[ \frac{1+\gamma}{\Gamma(2\gamma+1)} \right]^{1/2} ; \gamma = (1-\alpha^2 Z^2)^{1/2}; \lambda = \alpha Z m \quad (A.5)$$

We point out that the integral representation which occurs in eqs. A.3 and A.4 arises from the integral representation of the "negative power" term $r^{\gamma-1}$ [16]:

$$r^{\gamma-1} = \frac{\lambda^{1-\gamma}}{\Gamma(1-\gamma)} \int_0^\infty dx \, x^{-\gamma} e^{-\lambda r x} = \frac{\lambda^{1-\gamma}}{\Gamma(2-\gamma)} \int_0^\infty dx \, x^{1-\gamma} \frac{d}{dx}(e^{-\lambda r x}) \quad (A.6)$$

It is well known that the "small" components of the bispinor (A.2) are at least of the order $\beta^2 = \left(\frac{v}{c}\right)^2 \sim (\alpha Z)^2$ and can be ommited when we put $\gamma=1$, so that in the nonrelativistic region the exact Dirac spinor may be approximated by the nonrelativistic ground state wave function.

From the integral representations A.3 and A.4 we observe that the approximation $\gamma=1$ is fairly good since the largest contribution to the integral is given by the vicinity of the origin. This is obviously true if the photon energy $\omega = |\vec{k}|$ is not too large. For a value of the energy $\omega > \lambda - \alpha Z m$, the variable term $\lambda^2(2x+x^2)$ in the denominator $\left[(\vec{p}-\vec{k})^2 + \lambda^2(1+x)^2\right]^2$ becomes comparable with the constant term $(\vec{p}-\vec{k})^2 + \lambda^2$ only for higher values of $x$. This means that the domain of values of $x$ which give relevant contributions to the integral must be extended beyond the vicinity of the origin and the nonrelativistic approach is no longer valid. This conclusion is in accordance with the fact that in the case of large photon energies and large momentum transfer $\Delta = \vec{k}_2 - \vec{k}_1$ the contribution to the Rayleigh amplitude is mainly due to the small values of $r$ in the coulombian spinor, thus the term $r^{\gamma-1}$ must be exactly considered.

All these statements may be sustained when we get the nonrelativistic expression of the atomic factor $\vartheta_{NR}$ (eq. 2.19). Indeed, if we calculate the relativistic form-factor given by eq. (2.17), we obtain

$$\vartheta_{REL} = \left(1 + \frac{\Delta^2}{4\lambda^2}\right)^{-\gamma} \frac{\sin\left(2\gamma \arctan\frac{\Delta}{2\lambda}\right)}{\gamma\frac{\Delta}{\lambda}} \tag{A.7}$$

where the contribution of the "small" component of the bispinor is $\sqrt{\frac{1-\gamma}{1+\gamma}}$ times the contribution of the "large" component, so that the neglect of the "small" component is a good approximation. Taking into account that $\Delta = \vec{k}_2 - \vec{k}_1 = 2\omega\sin\frac{\theta}{2}$, eq. (A.7) may be written in the form

$$\vartheta_{REL} = \frac{\gamma^{-1}}{\left(1 + \frac{\alpha^2 Z^2}{(1+\gamma)^2}\frac{\omega^2}{\omega_{th}^2}\sin^2\frac{\theta}{2}\right)^{1+\gamma}} \left[\cos\left(2(1-\gamma)\arctan\frac{\omega\sin\frac{\theta}{2}}{\alpha Zm}\right) - \sin\left(2(1-\gamma)\arctan\frac{\omega\sin\frac{\theta}{2}}{\alpha Zm}\right)\frac{\alpha^2 Z^2 m^2 - \omega^2\sin^2\frac{\theta}{2}}{2\alpha Zm\omega\sin\frac{\theta}{2}}\right] \tag{A.8}$$

The relativistic parameter $\gamma = (1 - \alpha^2 Z^2)^{1/2}$ is present in the denominator exponent due to the behavior near the origin of the coulombian Dirac spinor given by the factor $r^{\gamma-1}$. The nonrelativistic limit is simply obtained by taking $\gamma = 1$, which leads to a subevaluation of the contribution of the small distances to the Rayleigh matrix element. This underestimation of the matrix element is not important in the case of not too large momentum transfer. In the momentum representation, the eqs. (A.2) and (A.3) obviously show the same behavior: if we put $\gamma = 1$, the small component vanishes, while the large component turns into the nonrelativistic ground-state eigenfunction. Thus, if we put $\gamma = 1$ in eq. (A.8), we get the nonrelativistic expression of the atomic form factor, eq. (2.19):

$$\vartheta_{NR} = \frac{1}{\left(1 + \frac{\alpha^2 Z^2}{4}\frac{\omega^2}{\omega_{th}^2}\sin^2\frac{\theta}{2}\right)^2} \tag{A.9}$$

We would like to emphasize that in the case of the amplitudes $\Pi_{jk}(\Omega)$, the most rough approximation occurs when we put $\gamma = 1$ in the exponent of the eqs. (A3) and (A4).

It has been proved by Hostler [9] that the iterated relativistic Coulombian Green function including the first two terms is

$$G_I(\vec{r}_2, \vec{r}_1; \Omega) = \left[I - \frac{1}{2\Omega}\vec{\alpha}\left(\vec{P}_2 + \vec{P}_1\right)\right] G_0(\vec{r}_2, \vec{r}_1; \Omega) \tag{A.10}$$

which in the momentum space can be written as

$$G_I(\vec{p}_2, \vec{p}_1; \Omega) = \left[I - \frac{1}{2\Omega}\vec{\alpha}\left(\vec{p}_2 - \vec{p}_1\right)\right] G_0(\vec{p}_2, \vec{p}_1; \Omega) \tag{A.11}$$

In the nonrelativistic region the operator $\frac{1}{2\Omega}\vec{\alpha}\left(\vec{P}_2 + \vec{P}_1\right)$ which gives the first iteration should be replaced by $\frac{1}{2m\Omega}\left(\vec{P}_2^2 + \vec{P}_1^2\right)$ so that the eq. (A.8) becomes

$$G_I(\vec{p}_2, \vec{p}_1; \Omega) = \left[1 - \frac{1}{2m\Omega}\left(\vec{p}_2^{\,2} + \vec{p}_1^{\,2}\right)\right] G_0(\vec{p}_2, \vec{p}_1; \Omega) \tag{A.12}$$

The neglect of the first iteration may be justified with the same arguments as in the case of the small components of the Dirac spinor. Moreover, as Hostler has shown, ( [9] eq. 3.16 and the comments below) that a consistent calculation requires that the first iteration to the coulombian Green function and the small components of the Dirac spinor have to be either taken into consideration or neglected together.

______________________________________________